\documentclass[12pt]{article}

\usepackage{epsfig}
\usepackage{amssymb}
\usepackage{amsmath}
\usepackage{amsfonts}

\newcommand{\R}{\mathbb{R}}
\newcommand{\C}{\mathbb{C}}
\newcommand{\Z}{\mathbb{Z}}

\newcommand{\fa}{\mathfrak{a}}
\newcommand{\fb}{\mathfrak{b}}
\newcommand{\fc}{\mathfrak{c}}
\newcommand{\fd}{\mathfrak{d}}
\newcommand{\fm}{\mathfrak{m}}
\newcommand{\fn}{\mathfrak{n}}

\newcommand{\fq}{\mathfrak{q}}

\newcommand{\cP}{\mathcal{P}}
\newcommand{\cQ}{\mathcal{Q}}

\newcommand{\cT}{\mathcal{T}}

\newcommand{\be}{\begin{equation}}
\newcommand{\ee}{\end{equation}}
\newcommand{\bea}{\begin{eqnarray}}
\newcommand{\eea}{\end{eqnarray}}
\newcommand{\nn}{\nonumber}
\newcommand{\kt}{\rangle}
\newcommand{\br}{\langle}

\newcommand{\ed}{\end{document}}

\newcommand{\np}{\newpage}

\newcommand{\bi}{\begin{itemize}}
\newcommand{\ei}{\end{itemize}}

\newcommand{\xtau}{{\mbox{\large{$\Theta$}}}}
\begin{document}

\title{$\cQ\cT$-Symmetry and Weak Pseudo-Hermiticity}
\author{\\
Ali Mostafazadeh
\\
\\
Department of Mathematics, Ko\c{c} University,\\
34450 Sariyer, Istanbul, Turkey\\ amostafazadeh@ku.edu.tr}
\date{ }
\maketitle

\begin{abstract}

For an invertible (bounded) linear operator ${\cal Q}$ acting in a
Hilbert space ${\cal H}$, we consider the consequences of the
${\cal QT}$-symmetry of a non-Hermitian Hamiltonian $H:{\cal
H}\to{\cal H}$ where ${\cal T}$ is the time-reversal operator. If
$H$ is symmetric in the sense that ${\cal T}H^\dagger{\cal T}=H$,
then ${\cal QT}$-symmetry is equivalent to ${\cal
Q}^{-1}$-weak-pseudo-Hermiticity. But in general this equivalence
does not hold. We show this using some specific examples. Among
these is a large class of non-${\cal PT}$-symmetric Hamiltonians
that share the spectral properties of ${\cal PT}$-symmetric
Hamiltonians.

\vspace{5mm}

\vspace{2mm}


\end{abstract}

%\tableofcontents
%\textheight = 22cm \topskip = -1cm \topmargin = -1cm

\section{Introduction}

Among the motivations for the study of the ${\cal PT}$-symmetric
quantum mechanics is the argument that \emph{${\cal PT}$-symmetry
is a more physical condition than Hermiticity because ${\cal
PT}$-symmetry refers to ``space-time reflection symmetry'' whereas
Hermiticity is ``a mathematical condition whose physical basis is
somewhat remote and obscure'' \cite{bender-ajp}.} This statement
is based on the assumption that the operators ${\cal P}$ and
${\cal T}$ continue to keep their standard meanings, as parity
(space)-reflection and time-reversal operators, also in ${\cal
PT}$-symmetric quantum mechanics. But this assumption in not
generally true, for unlike ${\cal T}$ the parity operator ${\cal
P}$ loses its connection to physical space once one endows the
Hilbert space with an appropriate inner product to reinstate
unitarity. This is because for a general ${\cal PT}$-symmetric
Hamiltonian, such as $H=p^2+x^2+ix^3$, the $x$-operator is no
longer a physical observable, the kets $|x\kt$ do not correspond
to localized states in space, and ${\cal P}:=\int_{-\infty}^\infty
dx\,|x\kt\br -x|$ does not mean space-reflection
\cite{jpa-2004b,jpa-2005b}.\footnote{The space reflection operator
is given by $\int_{-\infty}^\infty dx\,|\xi^{(x)}\kt\br
\xi^{(-x)}|$ where $|\xi^{(x)}\kt$ denote the (localized)
eigenkets of the pseudo-Hermitian position operator $X$,
\cite{jpa-2004b}.} Furthermore, it turns out that one cannot
actually avoid using the mathematical operations such as Hermitian
conjugation ($A\to A^\dagger$)~\footnote{The adjoint $A^\dagger$
of an operator $A:{\cal H}\to{\cal H}$ is defined by the condition
$\br\psi|A\phi\kt=\br A^\dagger|\phi\kt$ where $\br\cdot|\cdot\kt$
is the defining inner product of the Hilbert space ${\cal H}$.} or
transposition ($A\to A^t:={\cal T}A^\dagger{\cal T}$) in defining
the notion of an observable in ${\cal PT}$-symmetric quantum
mechanics \cite{comment,bender-erratum}.

What makes ${\cal PT}$-symmetry interesting is not its physical
appeal but the fact that ${\cal PT}$ is an antilinear
operator.\footnote{This means that ${\cal
PT}(a_1\psi_1+a_2\psi_2)=a_1^*{\cal PT}\psi_1+a_2^*{\cal
PT}\psi_2$, where $a_1,a_2$ are complex numbers and
$\psi_1,\psi_2$ are state vectors.} In fact, the spectral
properties of ${\cal PT}$-symmetric Hamiltonians
\cite{bender-prl-1998} that have made them a focus of recent
interest follow from this property. In general, if a linear
operator $H$ commutes with an antilinear operator $\xtau$, the
spectrum of $H$ may be shown to be pseudo-real, i.e., as a subset
of complex plane it is symmetric about the real axis. In
particular, nonreal eigenvalues of $H$ come in complex-conjugate
pairs. If $H$ is a diagonalizable operator with a discrete
spectrum the latter condition is necessary and sufficient for the
pseudo-Hermiticity of $H$ \cite{p3}.

In \cite{jpa-2006b}, we showed that the spectrum of the
Hamiltonian $H=p^2+z\delta(x)$ is real and that one can apply the
methods of pseudo-Hermitian quantum mechanics \cite{jpa-2004b} to
identify $H$ with the Hamiltonian of a unitary quantum system
provided that the real part of $z$ does not
vanish.\footnote{Otherwise $H$ has a spectral singularity and it
cannot define a unitary time-evolution regardless of the choice of
the inner product.} This Hamiltonian is manifestly non-${\cal
PT}$-symmetric. The purpose of this paper is to offer other
classes of non-${\cal PT}$-symmetric Hamiltonians that enjoy the
same spectral properties.

In the following, we shall use ${\cal H}$ and ${\cal T}$ to denote
a (separable) Hilbert space and an invertible antilinear operator
acting in ${\cal H}$, respectively. For ${\cal H}=L^2(\R^d)$, we
define ${\cal T}$ by \cite{isham}
    \be
    (\cT\psi)(\vec x):=\psi(\vec x)^*,
    \label{T-def1}
    \ee
for all $\psi\in L^2(\R^d)$ and $\vec x\in\R^d$. For ${\cal
H}=\C^N$, we identify it with complex-conjugation: For all $\vec
z\in \C^N$,
    \be
    \cT\vec z:=\vec{z}\,^*.
    \label{T-def2}
    \ee

\section{${\cal QT}$-Symmetry}

Consider a Hamiltonian operator $H$ acting in ${\cal H}$ and
commuting with an arbitrary invertible antilinear operator
$\xtau$. Because ${\cal T}$ is also invertible and antilinear, we
can express $\xtau$ as $\xtau={\cal QT}$ where ${\cal
Q}:=\xtau{\cal T}$ is an invertible linear operator. This suggests
the investigation of ${\cal QT}$-symmetric Hamiltonians $H$,
    \be
    [H,{\cal QT}]=0,
    \label{comm}
    \ee
where ${\cal Q}$ is an invertible linear operator. Note that
${\cal Q}$ need not be a Hermitian operator or an involution,
i.e., in general ${\cal Q}^\dagger\neq{\cal Q}$ and ${\cal
Q}^2\neq 1$.

We can easily rewrite~(\ref{comm}) in the form
    \be
    {\cal T}H{\cal T}={\cal Q}^{-1}H{\cal Q},
    \label{AT1}
    \ee
This is similar to the condition that $H$ is ${\cal
Q}^{-1}$-weakly-pseudo-Hermitian
\cite{solombrino,bagchi-quesne,znojil,jmp-2006b}:
    \be
    H^\dagger={\cal Q}^{-1}H{\cal Q}.
    \label{wph}
    \ee
In fact, (\ref{AT1}) and (\ref{wph}) coincide if and only if
    \be
    {\cal T}H^\dagger{\cal T}=H.
    \label{equi}
    \ee
The left-hand side of this relation is the usual ``transpose'' of
$H$ that we denote by $H^t$. Therefore, ${\cal QT}$-symmetry is
equivalent to ${\cal Q}^{-1}$-weak-pseudo-Hermiticity if and only
if $H^t=H$, i.e., $H$ is symmetric.\footnote{It is a common
practice to identify operators with matrices and define the
transpose of an operator $H$ as the operator whose matrix
representation is the transpose of the matrix representation of
$H$. Because one must use a basis to determine the matrix
representation, unlike $H^t:={\cal T}H^\dagger{\cal T}$, this
definition of transpose is basis-dependent. Note however that
$H^t$ agrees with this definition if one uses the position basis
$\{|\vec x\kt\}$ in $L^2(\R^d)$ and the standard basis in $\C^N$.}
For example, let $\vec a$ and $v$ be respectively complex vector
and scalar potentials, $\vec x\in\R^d$, and $d\in\Z^+$. Then the
Hamiltonian\footnote{Non-Hermitian Hamiltonians of this form have
been used in modelling localization effects in condensed matter
physics \cite{HN}.}
    \be
    H=\frac{[\vec p-\vec a(\vec x)]^2}{2m}+v(\vec x),
    \label{st}
    \ee
is symmetric if and only if $\vec a=\vec 0$. Supposing that $\vec
a$ and $v$ are analytic functions, the ${\cal QT}$-symmetry of
(\ref{st}), i.e., (\ref{AT1}) is equivalent to
    \be
    \frac{[\vec p+\vec a(\vec x)^*]^2}{2m}+v(\vec x)^*=
    \frac{[\vec p_{\cal Q}-\vec a(\vec x_{\cal Q})]^2}{2m}+
    v(\vec x_{\cal Q}),
    \label{AP2}
    \ee
where for any linear operator $L:{\cal H}\to{\cal H}$, we have
$L_{\cal Q}:={\cal Q}^{-1}L{\cal Q}$. Similarly the ${\cal
Q}^{-1}$-weak-pseudo-Hermiticity  of $H$, i.e., (\ref{wph}) means
    \be
    \frac{[\vec p-\vec a(\vec x)^*]^2}{2m}+v(\vec x)^*=
    \frac{[\vec p_{\cal Q}-\vec a(\vec x_{\cal Q})]^2}{2m}+
    v(\vec x_{\cal Q}),
    \label{wph2}
    \ee
As seen from (\ref{AP2}) and (\ref{wph2}), there is a one-to-one
correspondence between ${\cal QT}$-symmetric and ${\cal
Q}^{-1}$-weak-pseudo-Hermitian Hamiltonians of the standard form
(\ref{st}), namely that given such a ${\cal QT}$-symmetric
Hamiltonian $H$ with vector and scalar potentials $v$ and $a$,
there is a ${\cal Q}^{-1}$-weak-pseudo-Hermitian Hamiltonian $H'$
with vector and scalar potentials $v'=v$ and $a'=ia$. Note however
that $H$ and $H'$ are not generally isospectral.

\section{A Class of Matrix Models}

Consider two-level matrix models defined on the Hilbert space
${\cal H}=\C^2$ endowed with the Euclidean inner product
$\br\cdot|\cdot\kt$. In the following we explore the
$\cQ\cT$-symmetry and ${\cal Q}^{-1}$-weak-pseudo-Hermiticity of a
general Hamiltonian $H=\left(\begin{array}{cc} \fa&\fb\\
\fc&\fd\end{array}\right)$  for
$\cQ=\left(\begin{array}{cc} 1&0\\
\fq& 1\end{array}\right)$, where $\fa,\fb,\fc,\fd,\fq\in\C$.

\subsection{$\cQ\cT$-symmetric Two-Level Systems}
Imposing the condition that $H$ is $\cQ\cT$-symmetric (i.e.,
Eq.~(\ref{AT1}) holds) restricts $\fq$ to real and imaginary
values, and leads to the following forms for the Hamiltonian.
    \begin{itemize}
    \item For real $\fq$:
    \be
    H=\left(\begin{array}{cc} a&0\\
    c & a\end{array}\right),~~~~~~a,c\in\R.
    \label{q-real}
    \ee
    In this case $H$ is a non-diagonalizable operator with
    a real spectrum consisting of $a$.
    \item For imaginary $\fq$ ($\fq=iq$ with $q\in\R-\{0\}$):
    \be
    H=\left(\begin{array}{cc}
    a-\frac{i}{2}\, b\, q &b\\
    c+\frac{i}{2}(a-d) q & d+\frac{i}{2}\, b\,q
    \end{array}\right),~~~~~~a,b,c,d\in\R.
    \label{q-imaginary}
    \ee
    In this case the eigenvalues of $H$ are given by $E_\pm=
    \frac{1}{2}[a+d\pm\sqrt{(a-d)^2-b(bq^2-4c)}]$. Therefore,
    for $(a-d)^2\geq b(bq^2-4c)$, $H$ is a diagonalizable operator
    with a real spectrum; and for $(a-d)^2<b(bq^2-4c)$, $H$ is
    diagonalizable but its spectrum consists of a pair of
    (complex-conjugate) non-real eigenvalues. Furthermore,
    the degeneracy condition: $(a-d)^2=b(bq^2-4c)$ marks an
    exceptional spectral point \cite{kato,heiss} where $H$
    becomes non-diagonalizable. In fact, for $a=d$ and $b=0$ this
    condition is satisfied and $H$ takes the form (\ref{q-real}).
    Therefore, (\ref{q-imaginary}) gives the general form of
    $\cQ\cT$-symmetric Hamiltonians provided that $q\in\R$.
    \end{itemize}

\subsection{${\cal Q}^{-1}$-weakly-pseudo-Hermitian Two-Level Systems}
Demanding that $H$ is ${\cal Q}^{-1}$-weakly-pseudo-Hermitian does
not pose any restriction on the value of $\fq$. It yields the
following forms for the Hamiltonian.

    \begin{itemize}
    \item For $\fq=0$:
    \be
    H=\left(\begin{array}{cc} a&b_1+ib_2\\
    b_1-ib_2& d\end{array}\right),~~~~~~a,b_1,b_2,d\in\R.
    \label{q-zero}
    \ee
    In this case $\cQ$ is the identity operator and $H=H^\dagger$.
    Therefore, $H$ is a diagonalizable operator with a real
    spectrum.
    \item For $\fq\neq 0$:
    \be
    H=\left(\begin{array}{cc} a_1+ia_2&-\frac{2ia_2}{\fq}\\
    \frac{2ia_2}{\fq^*}& a_1-ia_2\end{array}\right),~~~~~~
    a_1,a_2\in\R,~~\fq\in\C-\{0\}.
    \label{q-nonzero}
    \ee
    In this case the eigenvalues of $H$ are given by $E_\pm=
    a_1\pm |a_2| |\fq|^{-1}\sqrt{4-|\fq|^2}$. Therefore, for
    $|\fq|<2$, $H$ is a diagonalizable operator with a real
    spectrum; and for $|\fq|>2$, $H$ is
    diagonalizable but its spectrum consists of a pair of
    (complex-conjugate) non-real eigenvalues. Again the degenerate
    case: $|\fq|=2$ corresponds to an exceptional point where $H$
    becomes non-diagonalizable.
    \end{itemize}

Comparing (\ref{q-imaginary}) with (\ref{q-zero}) and
(\ref{q-nonzero}) we see that $\cQ\cT$-symmetry and ${\cal
Q}^{-1}$-weak-pseudo-Hermiticity are totally different conditions
on a general non-symmetric Hamiltonian.\footnote{Note that this is
not in conflict with the fact that in view of the spectral
theorems of \cite{p1-p4,solombrino,solombrino2} both of these
conditions imply pseudo-Hermiticity of the Hamiltonian albeit with
respect to a pseudo-metric operator that differs from $\cQ^{-1}$,
\cite{jmp-2006b}.} For a symmetric Hamiltonian, we can easily show
using (\ref{q-zero}) and (\ref{q-nonzero}) that $\fq$ is either
real or imaginary and that $H$ takes the form (\ref{q-imaginary}).
The converse is also true, i.e., any symmetric Hamiltonian of the
form (\ref{q-imaginary}) is either real (and hence Hermitian) or
has the form (\ref{q-nonzero}). In summary, $\cQ\cT$-symmetry and
${\cal Q}^{-1}$-weak-pseudo-Hermiticity coincide if and only if
the Hamiltonian is a symmetric matrix.

\section{Unitary $\cQ$ and a Class of non-$\cP\cT$-Symmetric
Hamiltonians with a Pseudo-Real Spectrum}

If $\cQ$ is a unitary operator, the
$\cQ^{-1}$-weak-pseudo-Hermiticity (\ref{wph}) of a Hamiltonian
$H$ implies its $\cQ$-weak-pseudo-Hermiticity, i.e.,
$H^\dagger=\cQ^{-1}H\cQ$. This together with (\ref{wph}) leads in
turn to
    \be
    [H,\cQ^2]=0,
    \label{2q-sym}
    \ee
i.e., $\cQ^2$ is a symmetry generator. In the following we examine
some simple unitary choices for $\cQ$ and determine the form of
the $\cQ^{-1}$-weak-pseudo-Hermitian and $\cQ\cT$-symmetric
standard Hamiltonians.

Consider a standard non-Hermitian Hamiltonian (\ref{st}) in one
dimension and let
    \be
    \cQ=e^{\frac{i\ell p}{\hbar}}
    \label{transl}
    \ee
for some $\ell\in\R^+$. Then introducing
    \[a_1:=\Re(a),~~~a_2:=\Im(a),~~~v_1:=\Re(v),~~~v_2:=\Im(v),\]
where $\Re$ and $\Im$ stand for the real and imaginary parts of
their argument, and using the identities
    \be
    \cQ^{-1}p\cQ=p,~~~~~~~~~\cQ^{-1}x\cQ=x-\ell,
    \label{id}
    \ee
we can express the condition of the
$\cQ^{-1}$-weak-pseudo-Hermiticity of $H$, namely (\ref{wph2}), in
the form
    \bea
    &&a_1(x-\ell)=a_1(x),~~~~~~a_2(x-\ell)=-a_2(x),
    \label{a-per}\\
    &&v_1(x-\ell)=v_1(x),~~~~~~v_2(x-\ell)=-v_2(x).
    \label{v-per}
    \eea
This means that the real part of the vector and scalar potential
are periodic functions with period $\ell$ while their imaginary
parts are antiperiodic with period $\ell$. This confirms
(\ref{2q-sym}), for $H$ is invariant under the translation, $x\to
x+2\ell$, generated by $\cQ^2$. We can express $a_1,v_1$ and
$a_2,v_2$ in terms of their Fourier series. These have
respectively the following forms
    \bea
    &&\mbox{$\ell$-periodic real parts}:~~~
    \sum_{n=0}^\infty \left[c_{1n}\cos\left(\frac{2n\pi
    x}{\ell}\right)+d_{1n}\sin\left(\frac{2n\pi
    x}{\ell}\right)\right],
    \label{series-1}\\
    &&\mbox{$\ell$-antiperiodic imaginary parts}:~~~
    \sum_{n=0}^\infty \left[c_{2n}\cos\left(\frac{(2n+1)\pi
    x}{\ell}\right)+d_{2n}\sin\left(\frac{(2n+1)\pi
    x}{\ell}\right)\right],\quad\quad
    \label{series-2}
    \eea
where $c_{kn}$ and $d_{kn}$ are real constants for all
$k\in\{1,2\}$ and $n\in\{0,1,2,\cdots\}$.

Conversely if the real and imaginary parts of both the vector and
scalar potential have respectively the form (\ref{series-1}) and
(\ref{series-2}), the Hamiltonian is
$\cQ^{-1}$-weak-pseudo-Hermitian. In particular its spectrum is
pseudo-real; its complex eigenvalues come in complex-conjugate
pairs. These Hamiltonians that are generally non-${\cal
PT}$-symmetric acquire $\cQ\cT$-symmetry provided that they are
symmetric, i.e., $a_1=a_2=0$. A simple example is
    \[H=\frac{p^2}{2m}+\lambda_1\sin(2kx)+i\lambda_2\cos(5kx),\]
where $\lambda_1,\lambda_2\in\R$ and $k:=\ell^{-1}\in\R^+$.

Next, we examine the condition of $\cQ\cT$-symmetry of $H$, i.e.,
(\ref{AP2}). In view of (\ref{id}), this condition is equivalent
to (\ref{v-per}) and
    \bea
    &&a_1(x-\ell)=-a_1(x),~~~~~~a_2(x-\ell)=a_2(x),
    \label{a-per-2}
    \eea
which replaces (\ref{a-per}). Therefore $v$ has the same form as
for the case of a $\cQ^{-1}$-weak-pseudo-Hermitian Hamiltonian but
$a$ has $\ell$-antiperiodic real and $\ell$-periodic imaginary
parts. In particular, the Fourier series for real and imaginary
parts of $a$ have respectively the form (\ref{series-2}) and
(\ref{series-1}).

We again see that general $\cQ\cT$-symmetric Hamiltonians of the
standard form (\ref{st}) are invariant under the translation $x\to
x+2\ell$. This is indeed to be expected, because in view of
$[\cQ,\cT]=0$ we can express (\ref{AT1}) in the form
    \be
    H=\cQ^{-1}\cT H\cT\cQ
    \label{id-2}
    \ee
and use this identity to establish
    \[\cQ^2 H=\cQ\cT H\cT\cQ= \cQ\cT (\cQ^{-1}\cT H\cT\cQ)\cT\cQ=
    H\cQ^2.\]

The results obtained in this section admit a direct generalization
to higher-dimensional standard Hamiltonians. This involves
identifying $\cQ$ with a translation operator of the form
$e^{\frac{i\vec \ell\cdot\vec p}{\hbar}}$ for some $\vec
\ell\in\R^3-\{\vec 0\}$. It yields $\cQ\cT$-symmetric and
$\cQ^{-1}$-weakly-pseudo-Hermitian Hamiltonians with a pseudo-real
spectrum that are invariant under the translation $\vec x\to\vec
x-2\vec \ell$.

An alternative generalization of the results of this section to
(two and) three dimensions is to identify $\cQ$ with a rotation
operator:
    \be
    \cQ=e^{\frac{i\varphi\hat n\cdot\vec J}{\hbar}},
    \label{rot}
    \ee
where $\varphi\in(0,2\pi)$, $\hat n$ is a unit vector in $\R^3$,
and $\vec J$ is the angular momentum operator. Again $[\cQ,\cT]=0$
and we obtain generally non-$\cP\cT$-symmetric,
$\cQ^{-1}$-weak-pseudo-Hermitian and $\cQ\cT$-symmetric
Hamiltonians with a pseudo-real spectrum that are invariant under
rotations by an angle $2\varphi$ about the axis defined by $\hat
n$.

Choosing a cylindrical coordinate system whose $z$-axis is along
$\hat n$, we can obtain the general form of such standard
Hamiltonians.

The $\cQ^{-1}$-weak-pseudo-Hermiticity of $H$ implies that the
real and imaginary parts of the vector and scalar potentials (that
we identify with labels 1 and 2 respectively) satisfy
    \bea
    &&\vec a_1(\rho,\theta-\varphi,z)=\vec
    a_1(\rho,\theta,z),~~~~~
    \vec a_2(\rho,\theta-\varphi,z)=- \vec a_2(\rho,\theta,z),
    \label{rot-a}\\
    && v_1(\rho,\theta-\varphi,z)=v_1(\rho,\theta,z),~~~~~
    v_2(\rho,\theta-\varphi,z)=-v_1(\rho,\theta,z),
    \label{rot-v}
    \eea
where $(\rho,\theta,z)$ stand for cylindrical coordinates.
Similarly, the $\cQ\cT$-symmetry yields (\ref{rot-v}) and
    \be
    \vec a_1(\rho,\theta-\varphi,z)=-\vec
    a_1(\rho,\theta,z),~~~~~
    \vec a_2(\rho,\theta-\varphi,z)=\vec a_2(\rho,\theta,z).
    \label{rot-3}
    \ee
Again we can derive the general form of the Fourier series for
these potentials. Here we suffice to give the form of the general
symmetric Hamiltonian:
    \bea
    H&=&\frac{\vec p^2}{2m}+\sum_{n=0}^\infty\left[
    e_n(\rho,z)\cos(2n\omega\theta)+
    f_n(\rho,z)\sin(2n\omega\theta)+\right.\nn\\
    &&\hspace{2cm}\left. i\left\{
    g_n(\rho,z)\cos[(2n+1)\omega\theta]+
    h_n(\rho,z)\sin[(2n+1)\omega\theta]\right\}\right],
    \label{ex-3}
    \eea
where $e_n$, $f_n$, $g_n$, and $h_n$ are real-valued functions and
$\omega:=\varphi^{-1}\in\R^+$.

\section{A $\cQ\cT$-Symmetric and non-$\cP\cT$-Symmetric
Hamiltonian with a Real Spectrum}

In the preceding section we examined $\cQ\cT$-symmetric
Hamiltonians with a unitary $\cQ$. Because $\cP$ is also a unitary
operator, $\cQ\cT$-symmetry with unitary $\cQ$ may be considered
as a less drastic generalization of $\cP\cT$-symmetry. In this
section we explore a $\cQ\cT$-symmetric model with a non-unitary
$\cQ$.

Let $a$ and $a^\dagger$ be the bosonic annihilation and creation
operators acting in $L^2(\R)$ and satisfying $[a,a^\dagger]=1$,
$\fq\in\C-\{0\}$, and\footnote{The $\cQ$ considered in Section~3
may be viewed as a fermionic analog of (\ref{ee2}).}
    \be
    \cQ:=e^{\fq a}.
    \label{ee2}
    \ee
Consider the Hamiltonian operator
    \be
    H=\alpha\,a^2+\beta\,{a^\dagger}^2+\gamma\,\{a,a^\dagger\}+
    \fm\,a+\fn\,a^\dagger,
    \label{ee3}
    \ee
where $\alpha,\beta,\gamma,\fm,\fn\in\C$, and demand that $H$ be
$\cQ\cT$-symmetric. Inserting (\ref{ee3}) in (\ref{AT1}) and using
(\ref{T-def1}) and the identity
$\cQ^{-1}a^\dagger\cQ=a^\dagger-\fq$, we obtain
    \[
    \alpha^*=\alpha,~~~\beta^*=\beta,~~~\gamma^*=\gamma,~~~
    \fm^*=\fm-2\gamma \fq,~~~\fn^*=\fn-2\beta \fq,~~~\fn \fq=0.\]
In particular, because $\fq\neq 0$, we have $\fn=0$ which in turn
implies $\beta=0$. Furthermore, assuming that $\gamma\neq 0$, we
find that $\fq$ must be purely imaginary, $\fq=iq$ with
$q\in\R-\{0\}$, and $\Im(\fm)=\gamma q$. In view of these
observations, $H$ takes the following simple form.
    \be
    H=\alpha\,a^2+\gamma\,\{a,a^\dagger\}+
    (\mu+i\gamma q)\,a,
    \label{ee4}
    \ee
where $\mu:=\Re(\fm)$, $\alpha,\mu\in\R$, and
$\gamma,q\in\R-\{0\}$.

Recalling that $\cP a\cP=-a$ and $\cT a\cT=a$, we see that for
$\mu\neq 0$, the $\cQ\cT$-symmetric Hamiltonian~(\ref{ee4}) is
non-$\cP\cT$-symmetric. We also expect that it must have a
pseudo-real spectrum. It turns out that actually the spectrum of
$H$ can be computed exactly.

To obtain the spectrum of $H$ we use its representation in the
basis consisting of the standard normalized eigenvectors $|n\kt$
of the number operator $a^\dagger a$. Using the following
well-known properties of $|n\kt$, \cite{sakurai},
    \[ a|n\kt=\sqrt
    n\:|n-1\kt,~~~~~~a^\dagger|n\kt=\sqrt{n+1}\:|n+1\kt,\]
we find for all $m,n\in\{0,1,2,\cdots\}$,
    \[ H_{mn}:=\br m|H|n\kt=\gamma(2n+1)\delta_{mn}+
    (\mu+i\gamma q)\sqrt{n}\:\delta_{m,n-1}+\alpha
    \sqrt{n(n-1)}\:\delta_{m,n-2}.\]
As seen from this relation the matrix $(H_{mn})$ is
upper-triangular with distinct diagonal entries and up to three
nonzero terms in each row. This implies that the eigenvalues $E_n$
of $(H_{mn})$ are identical with its diagonal entries, i.e.,
$E_n=\gamma(2n+1)$. In particular, $H$ has a discrete, equally
spaced, real spectrum that is positive for $\gamma>0$. In the
latter case $H$ is isospectral to a simple harmonic oscillator
Hamiltonian with ground state energy $\gamma$.

It is not difficult to see that for each $n\in\{0,1,2,\cdots\}$
the span of $\{|0\kt,|1\kt,\cdots,|n\kt\}$, which we denote by
${\cal H}_n$, is an invariant subspace of $H$. This observation
allows for the construction of a complete set of eigenvectors of
$H$ and establishes the fact that $H$ is a diagonalizable operator
with a discrete real spectrum. Therefore, in view of a theorem
proven in \cite{p2}, it is related to a Hermitian operator via a
similarity transformation, i.e., it is quasi-Hermitian
\cite{quasi}.

The existence of the invariant subspace ${\cal H}_n$ also implies
that the eigenvectors $|\psi_n\kt$ of $H$ corresponding to the
eigenvalue $E_n$ belong to ${\cal H}_n$, i.e., $|\psi_n\kt$ is a
linear combination of $|0\kt,|1\kt,\cdots,|n-1\kt$ and $|n\kt$.
This in turn allows for a calculation of $|\psi_n\kt$. For
example,
    \bea
    |\psi_0\kt&=&c_0|0\kt,~~~~~~~~~~
    |\psi_1\kt=c_1\left[|0\kt+\left(\frac{2\gamma}{\fm}\right)|1\kt
    \right],\nn\\
    |\psi_2\kt&=&c_2\left[|0\kt+\left(\frac{4\gamma\fm}{\fm^2+
    \alpha\gamma}\right)|1\kt+\left(\frac{2\sqrt2\gamma^2}{\fm^2+
    \alpha\gamma}\right)|2\kt\right],\nn
    \eea
where $c_0,c_1,c_2$ are arbitrary nonzero normalization constants
and $\fm=\mu+i\gamma q$.

\section{Concluding Remarks}

It is often stated that $\cP\cT$-symmetry is a special case of
pseudo-Hermiticity because $\cP\cT$-symmetric Hamiltonians are
manifestly $\cP$-pseudo-Hermitian. This reasoning is only valid
for symmetric Hamiltonians $H$ that satisfy $H^\dagger=\cT H\cT$.
In general to establish the claim that $\cP\cT$-symmetry is a
special case of pseudo-Hermiticity one needs to make use of the
spectral theorems of \cite{p1-p4,solombrino,solombrino2}. Indeed
what makes $\cP\cT$-symmetric Hamiltonians interesting is the
pseudo-reality of their spectrum. This is a general property of
all Hamiltonians that are weakly pseudo-Hermitian or possess a
symmetry that is generated by an invertible antilinear operator.
We call the latter $\cQ\cT$-symmetric.

In this article, we have examined in some detail the similarities
and differences between $\cQ\cT$-symmetry and
$\cQ^{-1}$-weak-pseudo-Hermiticity and obtained large classes of
symmetric as well as asymmetric non-$\cP\cT$-symmetric
Hamiltonians that share the spectral properties of the
$\cP\cT$-symmetric Hamiltonians. In particular, we considered the
case that $\cQ$ is a unitary operator and showed that in this case
$\cQ\cT$-symmetry and $\cQ^{-1}$-weak-pseudo-Hermiticity imply
$\cQ^2$-symmetry of the Hamiltonian. We also explored a concrete
example of a $\cQ\cT$-symmetric Hamiltonian with a non-unitary
$\cQ$ that is not $\cP\cT$-symmetric. We determined the spectrum
of this Hamiltonian, established its diagonalizability, and showed
that it is indeed quasi-Hermitian.

\np

%{\small

%}

\ed
\begin{thebibliography}{99}
\bibitem{bender-ajp} C.~M.~Bender, D.~C.~Brody and H.~F.~Jones,
Am.~J.~Phys.\ {\bf 71}, 1095 (2003).

\bibitem{jpa-2004b} A.~Mostafazadeh and A.\ Batal, J.~Phys.~A {\bf 37},
11645 (2004).

\bibitem{jpa-2005b} A.~Mostafazadeh, J.~Phys.~A {\bf 38}, 6557
(2005).

\bibitem{comment} A.~Mostafazadeh, Preprint: quant-ph/0407070.

\bibitem{bender-erratum} C.~M.~Bender, D.~C.~Brody and H.~F.~Jones, Phys.\ Rev.\
Lett.\ {\bf 92}, 119902 (2004).

\bibitem{bender-prl-1998} C.~M.~Bender and S.~Boettcher,
Phys.\ Rev.\ Lett.\ {\bf 80}, 5243 (1998).

\bibitem{p3} A.~Mostafazadeh, J.\ Math.\ Phys.\ {\bf 43}, 3944
(2002).

\bibitem{jpa-2006b} A.~Mostafazadeh, J.~Phys.~A {\bf 39}, 10171
(2006).

\bibitem{isham} C.~J.~Isham, {\em Lectures on Quantum Theory},
Imperial College Press, London, 1995.

\bibitem{kato} T.~Kato, {\em Perturbation Theory for Linear
Operators}, Springer, Berlin, 1995.

\bibitem{heiss} W.~D.~Heiss, J.~Phys.~A {\bf 37}, 2455 (2004).

\bibitem{solombrino} L.~Solombrino, J.\ Math.\ Phys.\ {\bf 43},
5439 (2002).

\bibitem{bagchi-quesne} B.~Bagchi and C.~Quesne, Phys.~Lett.~A
{\bf 301}, 173 (2002).

\bibitem{znojil} M.~Znojil, Phys.~Lett.~A {\bf 353}, 463 (2006).

\bibitem{jmp-2006b} A.~Mostafazadeh, J.\ Math.\ Phys.\ {\bf 47},
092101 (2006).

\bibitem{HN} N.~Hatano, D.~R.~Nelson, Phys.\ Rev.~B
{\bf 56}, 8651 (1997); ibid {\bf 58}, 8384 (1998).

\bibitem{p1-p4} A.~Mostafazadeh, J.\ Math.\ Phys.\ {\bf 43}, 205
(2002); and 6343 (2002); Erratum: ibid, {\bf 44}, 943 (2003).

\bibitem{solombrino2} G.~Scolarici and L.~Solombrino, J.~Math.\
Phys.~{\bf 44}, 4450 (2003).

\bibitem{sakurai} J.~J.~Sakurai, {\em Modern Quantum Mechanics,}
Addisson-Wesley, New York, 1994.

\bibitem{p2} A.~Mostafazadeh, J.\ Math.\ Phys.\ {\bf 43},
2814 (2002).

\bibitem{quasi} F.~G.~Scholtz, H.~B.~Geyer, and F.~J.~W.~Hahne,
Ann.\ Phys.\ (NY) {\bf 213} 74 (1992).

\end{thebibliography}
